\documentclass[aps,prd,twocolumn,preprintnumbers,superscriptaddress,dblfloatfix]{revtex4-1}
\usepackage{bbm}
\usepackage{mathrsfs}
\usepackage{slashed}
\usepackage{caption}
\usepackage{epstopdf}
\usepackage[normalem]{ulem}
\usepackage[bottom]{footmisc}
\usepackage{subcaption}
\usepackage{bbold}
\usepackage{threeparttable}
\usepackage{booktabs}
\usepackage{changepage}
\usepackage[utf8]{inputenc}

\usepackage{grffile}

\usepackage{graphicx}  % needed for figures
\usepackage{dcolumn}   % needed for some tables
\usepackage{bm}        % for math
\usepackage{amssymb}   % for math
\usepackage{setspace}
\usepackage{amsmath, amssymb, setspace}
\usepackage{array}
\usepackage{booktabs}
\usepackage{caption}
\usepackage{indentfirst}
\usepackage{float}
\usepackage{lmodern}
\usepackage{multirow}
\usepackage{soul}
\usepackage[normalem]{ulem}

\usepackage{braket}
\usepackage{comment}

\usepackage{adjustbox} 
\newcommand{\ER}{E_{\rm R}}

\newcommand{\DoBox}[1]{\begin{center}
\color{red}\fbox{
\begin{minipage}{0.9\textwidth}

\end{minipage}}
\end{center}}

\newcommand{\Eq}[1]{Eq.~(\ref{#1})}

\newcommand{\Tab}[1]{Table~\ref{#1}}

\usepackage{color}

\begin{document}

%\preprint{\%}

\title{Geoneutrinos in Large Direct Detection Experiments}

\author{Graciela B. Gelmini}
\email[]{gelmini@physics.ucla.edu}
\affiliation{Department of Physics and Astronomy, University of California, Los Angeles\\
Los Angeles, CA 90095-1547, USA}

\author{Volodymyr Takhistov}
\email[]{vtakhist@physics.ucla.edu}
\affiliation{Department of Physics and Astronomy, University of California, Los Angeles\\
Los Angeles, CA 90095-1547, USA}

\author{Samuel J. Witte}
\email[]{sam.witte@ific.uv.es}
\affiliation{Instituto de F\'{i}sica Corpuscular (IFIC), CSIC-Universitat de Val\`{e}ncia, Apartado de Correos 22085, E-46071 Valencia, Spain}

%%%%%%%%%%%%%%%%%%%%%%%%%%%

\date{\today}

\begin{abstract}
Geoneutrinos can provide a unique insight into Earth's interior, its central engine and its formation history. We study the detection of geoneutrinos in large direct detection experiments, which has been considered non-feasible.  We compute the geoneutrino-induced electron and nuclear recoil spectra in different materials, under several optimistic assumptions.  We identify germanium as the most promising target element due to the low nuclear recoil energy threshold that could be achieved. The minimum exposure required for detection would be $\mathcal{O}(10)$ tonne-years. The realistic low thresholds achievable in germanium and silicon permit the detection of $^{40}$K geoneutrinos. These are particularly important to determine Earth's formation history but they are below the kinematic threshold of inverse beta decay, the detection process used in scintillator-based experiments.  
\end{abstract}

\maketitle

\section{Introduction}

The nature of dark matter (DM) remains one of the most pressing issues in modern physics. While weakly interacting massive particles (WIMPs) remain a theoretically well-motivated DM candidate, despite significant efforts no convincing detection signatures have been observed and a sizable portion of the allowed parameter space has been constrained.~Planned next generation large scale direct detection experiments will further explore the uncharted corners of WIMP interaction type, candidate mass and cross section. The experiments will eventually encounter an irreducible background due to coherent neutrino-nuclear interactions of solar as well as other neutrinos, i.e. the ``neutrino floor'' (e.g.~\cite{Billard:2013qya,Monroe:2007xp,Gelmini:2018ogy}). The neutrino floor can obfuscate the potential DM signal. Hence, this could constitute a major obstacle in making definitive statements regarding DM observations. However, the neutrino signal, which experiments will surely observe, as well as possible other sources, constitute in themselves interesting targets of exploration. It is thus imperative and timely to fully explore the scientific scope of large direct detection experiments beyond their main scope as DM detectors.

The next generation  of large scale direct detection experiments constitute a promising venue as neutrino detectors.~This has been bolstered by the recent detection of the coherent  neutrino-nucleus scattering by the COHERENT experiment \cite{Akimov:2017ade}.~Both coherent neutrino-nucleus as well as elastic electron-neutrino scatterings in direct detection experiments have already been employed in a range of studies to assess future detection capabilities within the context of neutrino-related physics, including sterile neutrinos (e.g.~\cite{Pospelov:2011ha,Billard:2014yka}), non-standard neutrino interactions (e.g.~\cite{Harnik:2012ni,Dutta:2017nht}), supernovae explosions \cite{Chakraborty:2013zua}, solar neutrinos (e.g.~\cite{Billard:2014yka,Cerdeno:2017xxl,Essig:2018tss,Newstead:2018muu}) as well as DM annihilations~\cite{McKeen:2018pbb}.

Geoneutrinos~(see e.g.~\cite{Bellini:2013wsa} for review) are anti-neutrinos $\overline{\nu}_e$ that originate from radioactive decays of uranium ${}^{238}$U, thorium $^{232}$Th and potassium $^{40}$K within the Earth's interior.~They provide a unique insight into the Earth's composition, heat production mechanisms as well as its thermal evolution and formation. In particular, geoneutrinos can reveal crucial information about the Earth's energy budget, a fundamental quantity in geology.~They allow one to establish which fraction of energy originates from ongoing radiogenic decays of heat producing elements and which is primordial, associated with Earth's formation history.~Furthermore, geoneutrinos can test 
the hypothetical Earth's central reactor, proposed to source the Earth's magnetic field~\cite{Hollenbach:2001}. 

The study of geoneutrinos finally became a possibility with recent developments of neutrino detectors.~Geoneutrino observations have been reported by the KamLAND~\cite{Araki:2005qa} and Borexino experiments \cite{Bellini:2010hy}, through inverse beta decay (IBD) $\overline{\nu}_e + p \rightarrow e^+ + n$ interactions within scintillator materials.~Large uncertainties, originating from low statistics and high backgrounds, do not allow experiments to make  definitive statements regarding geophysics at the current stage. Further progress is expected with future scintillator-based experiments such as SNO+~\cite{Andringa:2015tza}, JUNO~\cite{An:2015jdp, Han:2015roa}, HANOHANO~\cite{Cicenas:2012cta} and Jinping~\cite{Beacom:2016nol,Wan:2016nhe,Sramek:2016}. Directional detection experiments that utilize electron-neutrino scattering have also been put forward \cite{Leyton:2017tza, Wang:2017etb}.

Following earlier work in the field (e.g. \cite{Monroe:2007xp}), a potential geoneutrino signal in direct detection experiments has been largely neglected, assumed to be too small for detection. Given the status of experimental searches, the plans for larger detectors as well as recent interest in geoneutrinos, it is timely to revisit this issue and reassess the geoneutrino observation potential of future large-scale direct detection experiments. This is the main purpose of this work. 

\section{Large Direct Detection Experiments}

A variety of proposals for the next generation multi-ton scale direct detection experiments have been put forth~\cite{Cushman:2013zza}. Due to the inherent uncertainty in the final designs, making definitive statements about their scientific capabilities requires making assumptions about their composition, energy threshold and backgrounds. We explore several of the most likely detector configurations, based on xenon(Xe), germanium(Ge), argon(Ar) and silicon(Si), assuming optimistic design realizations.

\begin{table}[tb]
	\begin{center}
		\vspace*{0cm}
		\begin{tabular}[c]{l|cccc} \hline\hline
			Target   & $A(Z)$ & Isotope & Energy & Max\\ 
			Material & & Fraction & Threshold  & Energy \\
			  & & & [eVnr] & [eVnr]  \\
\hline
		    xenon(Xe)  & 128(54)  & 0.019 & 100 & 267 \\
		               & 129(54)  & 0.264 &  &      \\
		               & 130(54)  & 0.041 &  &     \\
		               & 131(54)  & 0.212 &  &     \\
		               & 132(54)  & 0.269 &  &     \\
		               & 134(54)  & 0.104 &  &     \\
		               & 136(54)  & 0.089 &  &     \\
\hline
		 argon(Ar)     & 40(18)  & 0.996 & 100 & 855  \\
\hline
		germanium(Ge)  & 70(32)  & 0.208 & 40 & 489  \\
		               & 72(32)  & 0.275 &   &    \\
		               & 73(32)  & 0.077 &   &    \\
		               & 74(32)  & 0.363 &   &     \\
		               & 76(32)  & 0.076 &   &    \\
\hline
		  silicon(Si)  & 28(14)  & 0.922 & 78 & 1221 \\
		               & 29(14)  & 0.047 &   &   \\
		               & 30(14)  & 0.031 &   &   \\
		          
\hline\hline
		\end{tabular}
		\caption{Considered experimental configurations. The last column shows the maximum energy of our region of interest.}
		\label{tab:experiments}
		\vspace*{-0.5cm}
	\end{center}
\end{table}

In~\Tab{tab:experiments} we list the  experimental setups we consider, each with a different target element.~We do not include isotopic abundances of elements that contribute less than $1\%$. We optimistically assume that experiments have perfect detection efficiency and energy resolution. Furthermore, when considering neutrino coherent interactions with the nuclei the expected background is assumed to arise exclusively from neutrinos. This allows to treat all the analyses on the same footing and provide results independent of specific configurations that could change in the future. The same assumption is not realistic for neutrino-electron scattering, since the backgrounds there are expected to be significant. Thus, we do not analyze this channel.

The reactor neutrino background as well as the geoneutrino signal depend sensitively on the location of the experiment. We assume that the experiment is located in the Jinping Underground Laboratory in China, where the expected geoneutrino signal is the largest and the reactor background is small. Its depth (6720 m.w.e.) ensures that the background due to cosmogenic muons will be highly suppressed.

\begin{figure}[tb]
\centering
\includegraphics[width=.45\textwidth]{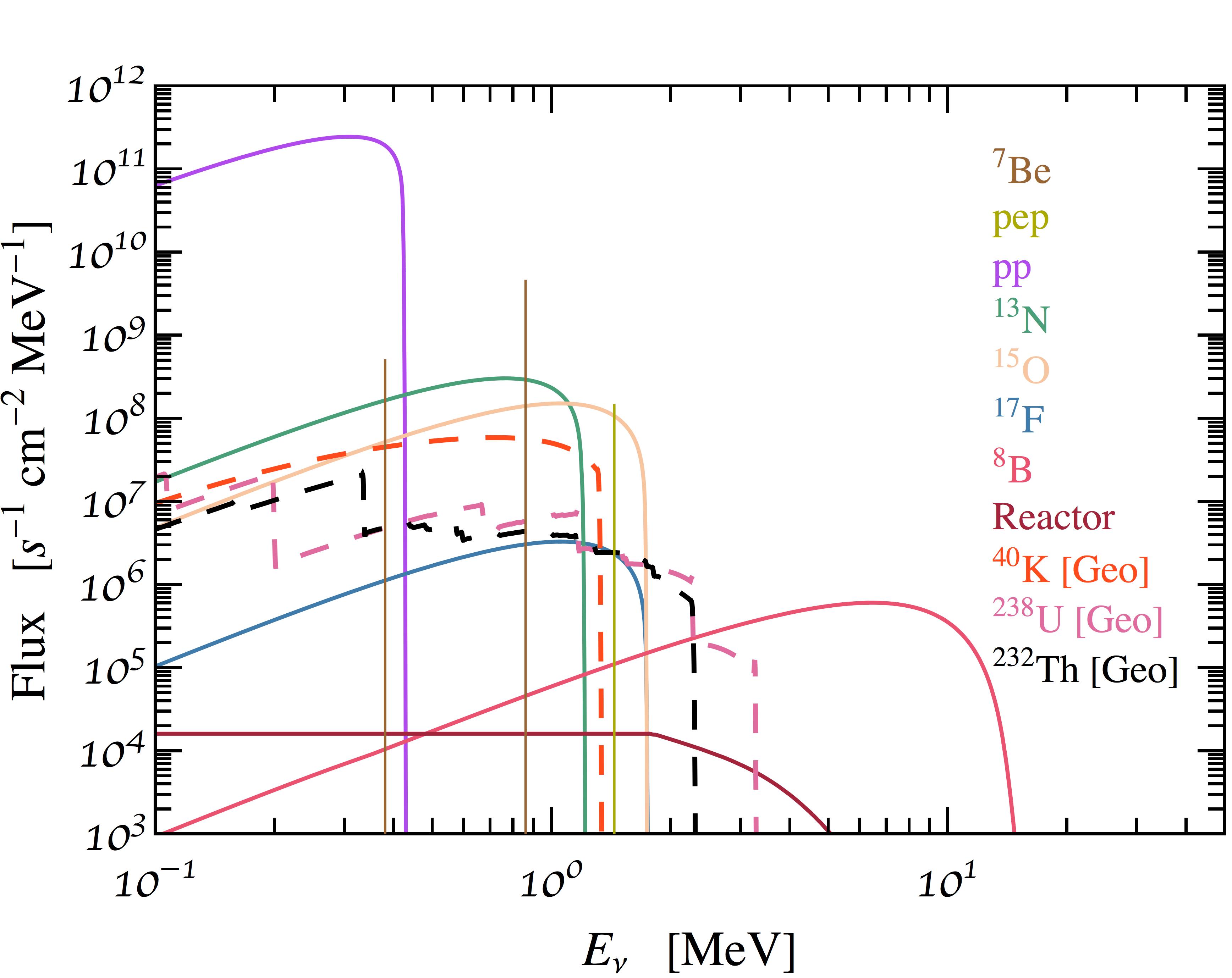}
\caption{\label{fig:neutrino_flux} Neutrino flux components for Jinping location. Geoneutrino signal as well as solar and reactor neutrino backgrounds are shown.}
\end{figure}

Recent analyses of xenon-based detectors, such as XENON1T~\cite{Aprile:2017iyp,Aprile:2018dbl} and LUX~\cite{Akerib:2016vxi}, set nuclear recoil thresholds of a few keV, when employing both scintillation and ionization signals.  These experiments can go to even lower recoil energies when using only their ionization signal, although their background rejection in this operation mode is less efficient.  In this way the XENON100 collaboration lowered its threshold to 0.7 keV for nuclear recoils~\cite{Aprile:2016wwo}. Thresholds as low as 0.1 keV have been used in some projections for future detectors, such as LZ~\cite{Akerib:2018lyp}. Similarly, the argon-based DarkSide-50 was  able to lower its threshold  for nuclear recoils to 0.6 keV with an analysis of only the ionization signal~\cite{Agnes:2018ves} (although their official analysis has some-what higher thresholds~\cite{Agnes:2018fwg}). Assuming optimistic realizations of future noble-gas based experiments, we use a threshold of 0.1 keV for xenon as well as argon.

Germanium and silicon will both be used in the future SuperCDMS SNOLAB experiment.~With their HV (high voltage) detectors SuperCDMS has the capacity to reach the nuclear recoil thresholds of 40 eV and 78 eV (see Table VII of Ref.~\cite{Agnese:2016cpb}). We adopt these values here. 

In order to maximize the geoneutrino signal when studying nuclear recoils, we will use data within a region of interest (ROI) defined by the recoil energies between the assumed experimental threshold energy and the maximum energy that can be imparted by any geoneutrino (see \Tab{tab:experiments}).

\section{Geoneutrino Signal and Background}

The geoneutrino signal as well as reactor and solar neutrino background components are displayed in Figure~\ref{fig:neutrino_flux}. We describe them in more detail in the section below.

\subsection{Geoneutrino signal}

\subsubsection{Spectrum}

\begin{figure}[tb]
\centering
\includegraphics[width=.45\textwidth]{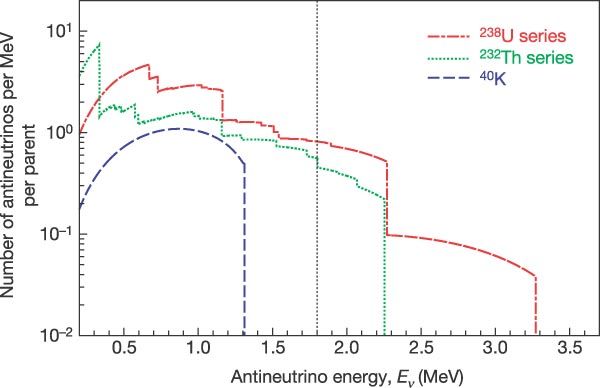}
\caption{\label{fig:geonuspec} The geoneutrino $\overline{\nu}_e$ spectrum expected from ${}^{238}$U, ${}^{232}$Th and ${}^{40}$K decay chains. Dotted vertical line at 1.8 MeV represents the kinematic threshold for IBD interactions. Reproduced from \cite{Araki:2005qa}.}
\end{figure}

The radioactive decays of  ${}^{238}$U, ${}^{232}$Th and ${}^{40}$K are responsible for more than 99\% of the Earth's radiogenic heat. These elements decay via a series of $\alpha$ and $\beta^-$ processes, which terminate at stable isotopes. Each $\beta^-$ decay results in the emission of $\overline{\nu}_e$. Comprehensive computations are needed to obtain the overall resulting $\overline{\nu}_e$ spectrum for each decay chain, including the spectral shapes and rates of individual decays for more than $80$ different branches in each chain \cite{Ludhova:2013hna}. In Fig.~\ref{fig:geonuspec} we display the resulting geoneutrino spectra for each element used in our study, as shown in Ref.~\cite{Araki:2005qa}. Note that the kinematic threshold of 1.8 MeV for inverse-beta decay (IBD) interactions makes detectors based on this reaction insensitive to geoneutrinos generated from the ${}^{40}$K decays. 

\subsubsection{Flux normalization}

The geoneutrino flux expected in different locations strongly depends on the amount and distribution of heat producing elements within Earth. The Earth's surface heat flow of $47\pm2$ TW \cite{davies:2010} is  composed partly of heat generated by radioactive decays in the crust and mantle and partly of heat from the secular planet cooling (i.e. primordial heat that is byproduct of the Earth's formation). While the crust contributions are relatively well known, the contributions from the mantle remain highly uncertain. The predicted radiogenic heat, originating from the heat producing elements, varies greatly between different bulk silicate Earth models \cite{Sramek:2012nk}, which do not include the Earth's metallic core.~The models can be generally classified as cosmochemical, geochemical and geodynamical, which make radiogenic heat predictions of $11 \pm 2$ TW, $17 \pm 3$ and $33 \pm 3$ TW, respectively.

To maximize the geoneutrino signal, we consider the most optimistic scenario of $100\%$ of Earth's heat being radiogenic, i.e. Earth's radiogenic heat comprising $\sim 47$~TW~\cite{Bellini:2013wsa}.~This choice is compatible with current geoneutrino measurements within few standard deviations \cite{Agostini:2015cba}. The location-dependent geoneutrino flux is particularly sensitive to the amount of crust beneath the laboratory site, which contains the largest portion of heat producing elements. Due to this reason, as well as substantial distance to the closest nuclear reactors, Jinping Underground Laboratory, situated near the Himalayan mountains in China where the amount of crust is maximal, has been previously identified as a favorable location for geoneutrino detection~\cite{Sramek:2016}. In principle, the geoneutrino flux is calculated by summing contributions from each Earth grid component, using a detailed geological map specifying heat producing element distribution and abundance. Assuming a nearly fully radiogenic Earth, we take the geoneutrino flux at Jinping (see Fig. 6 of Ref.~\cite{Wan:2016nhe}) corresponding to $\gtrsim 90$ TNU (terrestrial natural units), which is $\sim 2.5$ larger (see Fig. 2 of Ref.~\cite{Sramek:2016}) than the computed flux for each radiogenic component at the SNOLab (Sudbury, Canada) using the Reference Earth Model (see Table 2 of Ref.~\cite{huang:2013geomodel}). The geoneutrino fluxes we adopt are provided in \Tab{tab:geoflux}.~The corresponding flux uncertainties are taken from Ref.~\cite{huang:2013geomodel}.

\begin{table}[htb]
	\begin{center} \setlength{\extrarowheight}{2pt}
		\vspace*{0cm}
		\begin{tabular}[c]{l|cc} \hline\hline
			Neutrino   & Flux & Max Energy\\ 
			Component & $\mathrm{[cm^{-2} s^{-1}]}$ & $\mathrm{E_{\nu}~[MeV]}$
			\\\hline
			$\mathrm{Geo (\overline{\nu}_e, {}^{40}K)}$  & ~$5.47(1\pm0.006) \times 10^{7}$  & $1.32$     \\
			$\mathrm{Geo (\overline{\nu}_e, {}^{238}U)}$  & $1.22(1\pm0.012) \times 10^{7}$  & $3.99$     \\
			$\mathrm{Geo (\overline{\nu}_e, {}^{232}Th)}$  & $1.14(1\pm0.300) \times 10^{7}$  & $2.26$     \\
\hline\hline
		\end{tabular}
		\caption{Geoneutrino signal flux.}
		\label{tab:geoflux}
		\vspace*{-0.5cm}
	\end{center}
\end{table}

\subsection{Neutrino background}
\label{ss:nuback}

Since the geoneutrino signal ceases at neutrino energies above $\sim 5$ MeV, we shall only consider the backgrounds due to solar and reactor neutrinos (see Fig.~\ref{fig:neutrino_flux}), which dominate at energies $E_{\nu} \lesssim 20$ MeV. At higher neutrino energies, contributions from atmospheric and diffuse supernova neutrinos also become important. 

\begin{table}[t]
	\begin{center}  \setlength{\extrarowheight}{2pt}
		\vspace*{0cm}
		\begin{tabular}[c]{l|cc} \hline\hline
			Neutrino   & Flux & Max Energy\\ 
			Component & $\mathrm{[cm^{-2} s^{-1}]}$ & $\mathrm{E_{\nu}~[MeV]}$
			\\\hline
			$\mathrm{Solar (\nu_e, pp)}$  & ~$6.03(1\pm0.006) \times 10^{10}$  & $0.42$     \\
			$\mathrm{Solar (\nu_e, pep [line])}$  & $1.47(1\pm0.012) \times 10^{8}$  & $1.45$     \\
			$\mathrm{Solar (\nu_e, hep)}$  & $8.31(1\pm0.300) \times 10^{3}$  & $18.77$     \\
			$\mathrm{Solar (\nu_e, {}^{7}Be [line~1])}$  & $4.56(1\pm0.070) \times 10^{8}$  & $0.39$     \\
			$\mathrm{Solar (\nu_e, {}^{7}Be [line~2])}$  & $4.10(1\pm0.070) \times 10^{9}$  & $0.87$     \\
			$\mathrm{Solar (\nu_e, {}^{8}B)}$  & $4.59(1\pm0.140) \times 10^{6}$  & $16.80$     \\
		    $\mathrm{Solar (\nu_e, {}^{13}N)}$  & $2.17(1\pm0.140) \times 10^{8}$  & $1.20$ \\
		    $\mathrm{Solar (\nu_e, {}^{15}O)}$  & $1.56(1\pm0.150) \times 10^{8}$  & $1.73$ \\
		    $\mathrm{Solar (\nu_e, {}^{17}F)}$  & $3.40(1\pm0.170) \times 10^{6}$  & $1.74$ \\ 
\hline
		    $\mathrm{Reactor (\overline{\nu}_e})$  & $4.95 (1\pm0.100) \times 10^{4}$  & $10.00$ \\
\hline\hline
		\end{tabular}
		\caption{Solar and reactor neutrino background fluxes.}
		\label{tab:backflux}
		\vspace*{-0.5cm}
	\end{center}
\end{table}

\subsubsection{Solar neutrinos}

Solar electron neutrinos $\nu_e$ are produced as a byproduct of nuclear fusion reactions in the Sun (see  \cite{Robertson:2012ib} for review).
Multiple reaction chains contribute, varying in resulting neutrino flux and energy.
More than 98\% of the Sun's energy is produced through the proton-proton cycle. Starting with the initial reaction of $p + p \rightarrow ~^{2}{\rm H} + e^+ + \nu_e$, this cycle results in $pp$, $hep$, $pep$, $^{7}$Be and $^{8}$B neutrino components. The remaining solar energy is released through the Carbon-Nitrogen-Oxygen (CNO) cycle, which results in $^{13} $N, $^{15}$O and $^{17}$F neutrinos. 
The solar neutrino background is independent of the laboratory location.

Specific solar neutrino fluxes can be inferred from a solar model. The Standard Solar Model (GS98) \cite{Grevesse:1998bj} showed a good agreement with helioseismological studies.~On the other hand, more recent constructions (AGSS09) \cite{Asplund:2009fu} that posses a higher degree of internal consistency, appear to disagree with results from helioseismology.~This discrepancy is the so-called ``solar metallicity'' problem.~The CNO neutrinos are particularly promising in this context, given their ability to shed light on the Sun's internal composition, and their observation potential in future direct detection experiments has been recently explored~\cite{Billard:2014yka,Cerdeno:2017xxl,Newstead:2018muu}.
For our analysis we assume the solar flux normalizations and uncertainties as predicted by AGSS09 (see Table 2 of \cite{Robertson:2012ib}), which we specify in \Tab{tab:backflux}. The background in our study is dominated by $^8$B neutrinos. Since the $^8$B flux difference between the two solar models is only $\sim 20$\%, we do not expect a significant change in our sensitivity results if the alternative model is employed.

\subsubsection{Reactor neutrinos}

Reactor electron anti-neutrinos $\overline{\nu}_e$ (see \cite{Hayes:2016qnu} for a review) originate from fission $\beta$-decays of uranium ($^{235}$U and $^{238}$U) and plutonium ($^{239}$Pu and $^{241}$Pu) within the reactor fuel. Since the considered isotopes are short lived, the corresponding neutrino flux follows the reactor operation. The fraction of isotope contributions as well as the average energy release per fission are given in~\Tab{tab:fission}, following Ref.~\cite{Ma:2012bm}. As a reactor operates, its fuel composition and  relative isotope fractions change with time, in a manner specific for each individual reactor. We neglect these effects (an analysis including these contributions can be found in Ref.~\cite{Murayama:2000iq}).

\begin{table}[htb]
	\begin{center}
		\vspace*{0cm}
		\begin{tabular}[c]{c|cc} \hline\hline
			Isotope               & $f_k$& $E_k$ [MeV/fission]   \\\hline
			$\mathrm{{}^{235}U}$  & 0.58 & $202.36\pm0.26$     \\
			$\mathrm{{}^{238}U}$  & 0.07 & $205.99\pm0.52$     \\
			$\mathrm{{}^{239}Pu}$ & 0.30 & $211.12\pm0.34$     \\
			$\mathrm{{}^{241}Pu}$ & 0.05 & $214.26\pm0.33$     \\\hline\hline
		\end{tabular}
		\caption{Fission fraction and average released energy for each nuclear reactor isotope.}
		\label{tab:fission}
		\vspace*{-0.5cm}
	\end{center}
\end{table}

Unlike the flux of solar neutrinos, the reactor neutrino flux depends on the laboratory location and in particular on the characteristics and distances of the main contributing reactors~\cite{Baldoncini:2014vda}.~The flux of reactor anti-electron neutrinos from an isotope  $k$ is given by 
\begin{equation}
\phi_k(E) = \frac{R_{\bar{\nu}_e}}{4\pi d^2}\sum_k f_k S_k (E) \, ,
\end{equation}
where $R_{\bar{\nu}_e}$ is the emission rate, $d$ is the distance of a given reactor to the laboratory, $S_k(E)$ is the corresponding neutrino spectrum and $f_k$ is the fraction of the isotope $k$ in the reactor fuel. The neutrino emission rate is 
\begin{equation}
R_{\bar{\nu}_e} = N_{\nu, {\rm f}} \, \frac{P_{th}}{\sum_{k} f_k E_k} e \simeq 6 \times 10^{20} \Big(\dfrac{P_{th}}{\rm1~GW}\Big)e~{\rm s}^{-1}~,
\end{equation}
where $N_{\nu, {\rm f}} \simeq 6$ is the average number of anti-neutrinos produced per fission, $P_{th}$ is the thermal power output of the reactor, $E_k$ is the fission energy and $e = 0.75 \, (1 \pm 0.1)$ is the average reactor operation efficiency that includes shut-downs \cite{usenergy}. The uncertainty of the efficiency reflects an uncertainty of about $8\%$ in the reactor anti-neutrino spectrum (see Table 3 of Ref.~\cite{Kopeikin:2012zz}) as well as an uncertainty of few percent on the fission emission. This is the source of the reactor flux uncertainty in \Tab{tab:backflux}. Approximate analytic expressions for neutrino spectra $S_k(E_{\nu})$ have been constructed \cite{Vogel:1989iv,Mueller:2011nm,Huber:2011wv,Murayama:2000iq}. We adopt the spectra from the model of Ref.~\cite{Mueller:2011nm}, which is based on a phenomenological fit to data with an exponentiated polynomial. The resulting expression is 
\begin{equation}\label{eq:reactor_nuspec}
 S_k (E_\nu) = \dfrac{dN_{\nu}}{dE_{\nu}} = \exp \left(\sum_{i=1}^{6} \alpha_{i,k}E_\nu^{i-1} \right) ~,
\end{equation}
where $\alpha_{i,k}$ is the respective fit coefficient of order $i$. The best values of the fit coefficients, as obtained in Ref.~\cite{Mueller:2011nm}, are displayed in \Tab{tab:reactor_fit}. Strictly, these fits are only valid for $E_{\nu} \gtrsim 1.8$ MeV. For uranium, Ref.~\cite{Vogel:1989iv} does not find substantial deviations in the fit results for energies above $0.5$ MeV. For other elements, the flux at lower energies appears enhanced. Since this could be a model artifact and not of physical significance, we conservatively set $S_k (E_{\nu} < 1.8~\text{MeV}) = S_k (E_{\nu} = 1.8~\text{MeV})$ for all elements. This effectively appears as a ``kink'' feature visible in the resulting flux figures. Recent measurements of the reactor anti-neutrino spectra from Daya Bay~\cite{Adey:2018qct} indicate agreement with the predictions of the leading theoretical models of Huber-Muller~\cite{Huber:2011wv,Mueller:2011nm} and ILL-Vogel~\cite{Vogel:1980bk,Schreckenbach:1985ep,VonFeilitzsch:1982jw,Hahn:1989zr} at the level of $\sim10$\% difference on the predicted anti-neutrino yield. Since reactor background constitutes a sub-dominant contribution in our study, we do not expect reactor neutrino modeling uncertainty to significantly affect our results.

\begin{table}[tb]
  \setlength{\extrarowheight}{2pt}
  \setlength{\tabcolsep}{5pt}
  \begin{center}
	\begin{tabular}{c|c|c|c|c}  \hline\hline
	$i$  & $^{235}$ & $^{238}$U & $^{239}$Pu & $^{241}$Pu \\  
	\hline
	1 & 3.217 & 0.4833 & 6.413 & 3.251\\  \hline
	2 & -3.111 & 0.1927 & -7.432 & -3.204 \\  \hline
    3 & 1.395 & -0.1283 & 3.535 & 1.428 \\  \hline
    4 & -0.3690 & -0.006762 & -0.8820 & -0.3675 \\  \hline
    5 & 0.04445 & 0.002233 & 0.1025 & 0.04254 \\  \hline
	6 &  -0.002053 & -0.0001536 & -0.004550 & -0.001896 \\  \hline\hline
	\end{tabular}
  \end{center}
\caption{\label{tab:reactor_fit} Fitted values of the reactor neutrino spectrum $\alpha_{i, k}$ coefficients, for $i = 1$ to 6, for the dominant nuclear isotopes (denoted by $k$).}
\end{table}

We list the reactors considered in our study for the Jinping location in \Tab{tab:reactors}.  

\subsection{Recoil rates}

We consider two types of interactions in direct detection experiments, elastic electron neutrino scattering and coherent neutrino-nucleus scattering. 

For a neutrino flux $\phi_{\nu} (E_{\nu})$ the resulting differential event rate per unit time and detector mass as a function of the recoil energy $E_R$, per unit time and mass $m_I$ of a target nuclide $I$ in a detector is given by
\begin{equation} \label{eq:nu_diff_rate}
\dfrac{d R_{\nu}^I}{d E_R} = \dfrac{C_I}{m_I} \int_{E_{\nu}^{\rm min}} \phi_{\nu} (E_{\nu}) \dfrac{d \sigma^I (E_{\nu}, E_R)}{d E_R} d E_{\nu}~,
\end{equation}
where $d \sigma^I (E_{\nu}, E_R) / d E_R$ is the coherent neutrino-nucleus scattering differential cross-section and $C_I$ is the fraction of nuclide $I$ in the material. When several nuclides are present, for each type neutrino flux contribution $\phi_{\nu}(E_{\nu})$ the differential event rate is given by summing over them
\begin{equation} \label{eq:totnurate}
\dfrac{d R_{\nu}}{d E_R} = {\rm MT} \sum_I \dfrac{d R_{\nu}^I}{d E_R}~,
\end{equation}
where MT is the experimental exposure.

\begin{table}[tbp]
  \setlength{\extrarowheight}{2pt}
  \begin{center}
	\begin{tabular}{l|m{2.3cm}|c|c}  \hline \hline
	Reactor & Location &  Distance & Output \\
		 &  &  [km] & [MW$_t$] \\ \hline 
    Fangchengang  & 21$^\circ$40$'$15$''$N  108$^\circ$33$'$30$''$\,E   & 998 & 12110  \\ \hline
    Changjiang  & 19$^\circ$27$'$39$''$\,N 108$^\circ$53$'$60$''$\,E  & 1206 & 3806 \\ \hline
    Yangjiang  & 21$^\circ$42$'$30$''$\,N 112$^\circ$15$'$40$''$\,E  & 1239 & 17430 \\ \hline
    Taishan  & 21$^\circ$55$'$04$''$\,N 112$^\circ$58$'$55$''$\,E  & 1318 & 1980 \\ \hline
    Ling Ao  & 22$^\circ$36$'$17$''$\,N 114$^\circ$33$'$05$''$\,E  & 1422 & 11620 \\ \hline
    Fuqing  & 25$^\circ$26$'$45$''$\,N 119$^\circ$26$'$50$''$\,E  & 1763 & 17740 \\ \hline
    Hanbit  & 35$^\circ$24$'$54$''$\,N 126$^\circ$25$'$26$''$\,E  & 2463 & 16874 \\ \hline\hline
	\end{tabular}
  \end{center}
\caption{\label{tab:reactors}  List of selected most relevant nuclear reactors for Jinping. Columns contain, from left to right, the name of the reactor, the GPS location, the distance to the laboratory in kilometers and the thermal output of the reactor in MW.}  
		\vspace*{-0.5cm}
\end{table}

A similar expression holds for electrons if we neglect the effects of their atomic and molecular binding. In the approximation of considering electrons as free in the material, we need to take into account the charge number $Z_I$ in every nuclide $I$. Hence, the interaction rate in a detector becomes
\begin{equation} \label{eq:nu_diff_rate}
\dfrac{d R_{\nu}}{d E_R} = {\rm MT} \sum_I \dfrac{C_I}{m_I} Z_I \int_{E_{\nu}^{\rm min}} \phi_{\nu} (E_{\nu}) \dfrac{d \sigma^e (E_{\nu}, E_R)}{d E_R} d E_{\nu}~,
\end{equation}
where $d\sigma^e (E_{\nu}, E_R)/dE_R$ is the neutrino-electron elastic scattering differential cross-section.

For a target mass $m$ at rest ($m_I$ or the electron mass $m_e$), the minimum neutrino energy required to produce a recoil of energy $E_R$ is
\begin{equation}
E_{\nu}^{\rm min} = \sqrt{\dfrac{m E_{R}}{2}}~.
\end{equation}
The maximum recoil energy due to a collision with neutrino of energy $E_{\nu}$ is
\begin{equation}
E_R^{\rm max} = \dfrac{2 E_{\nu}^2}{m + 2 E_{\nu}}~.
\end{equation}

\begin{figure*}[tb]
        \centering
        \begin{subfigure}[b]{0.475\textwidth}
            \centering
            \includegraphics[width=\textwidth]{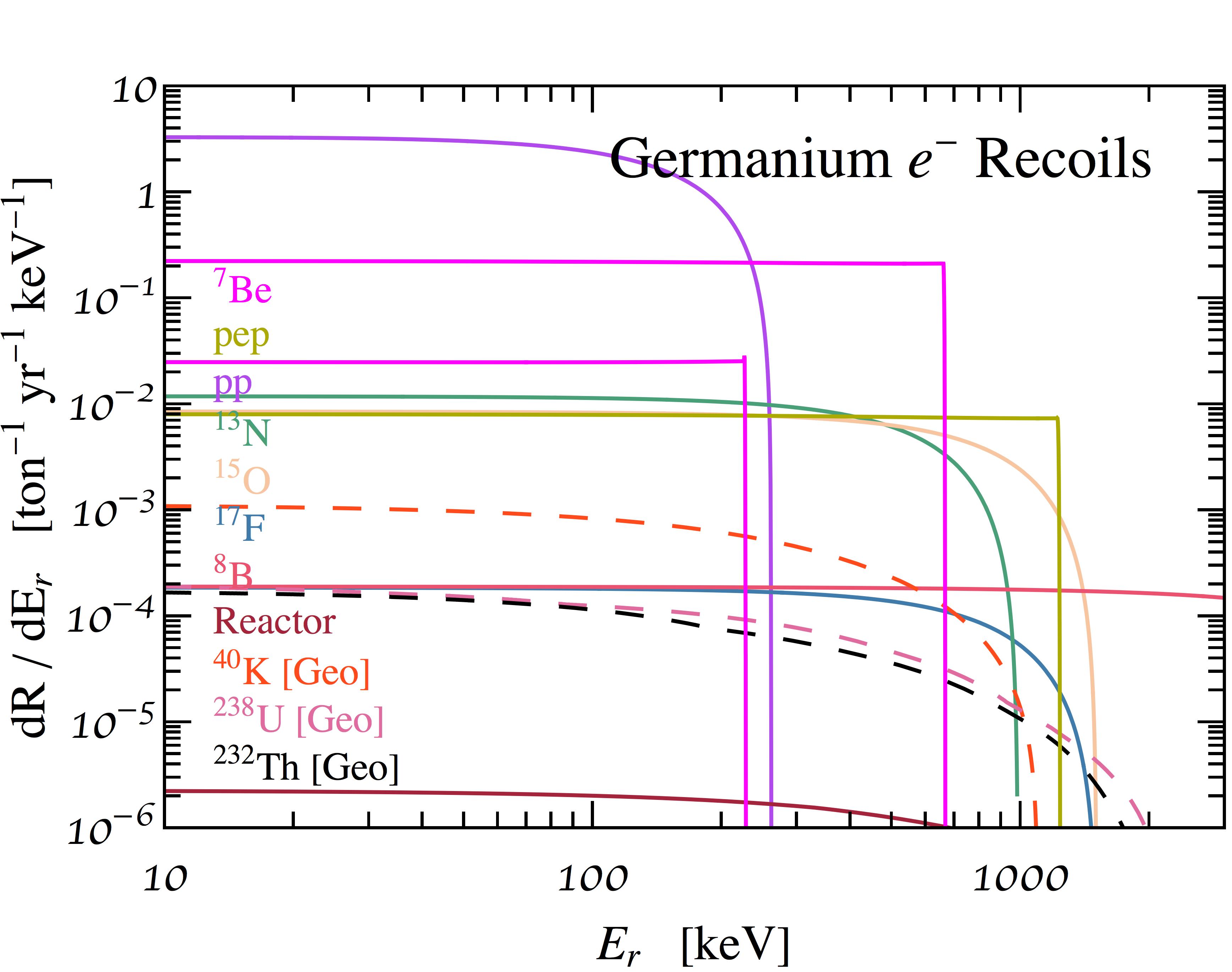} 
        \end{subfigure}
        \hfill
        \begin{subfigure}[b]{0.475\textwidth}  
            \centering 
            \includegraphics[width=\textwidth]{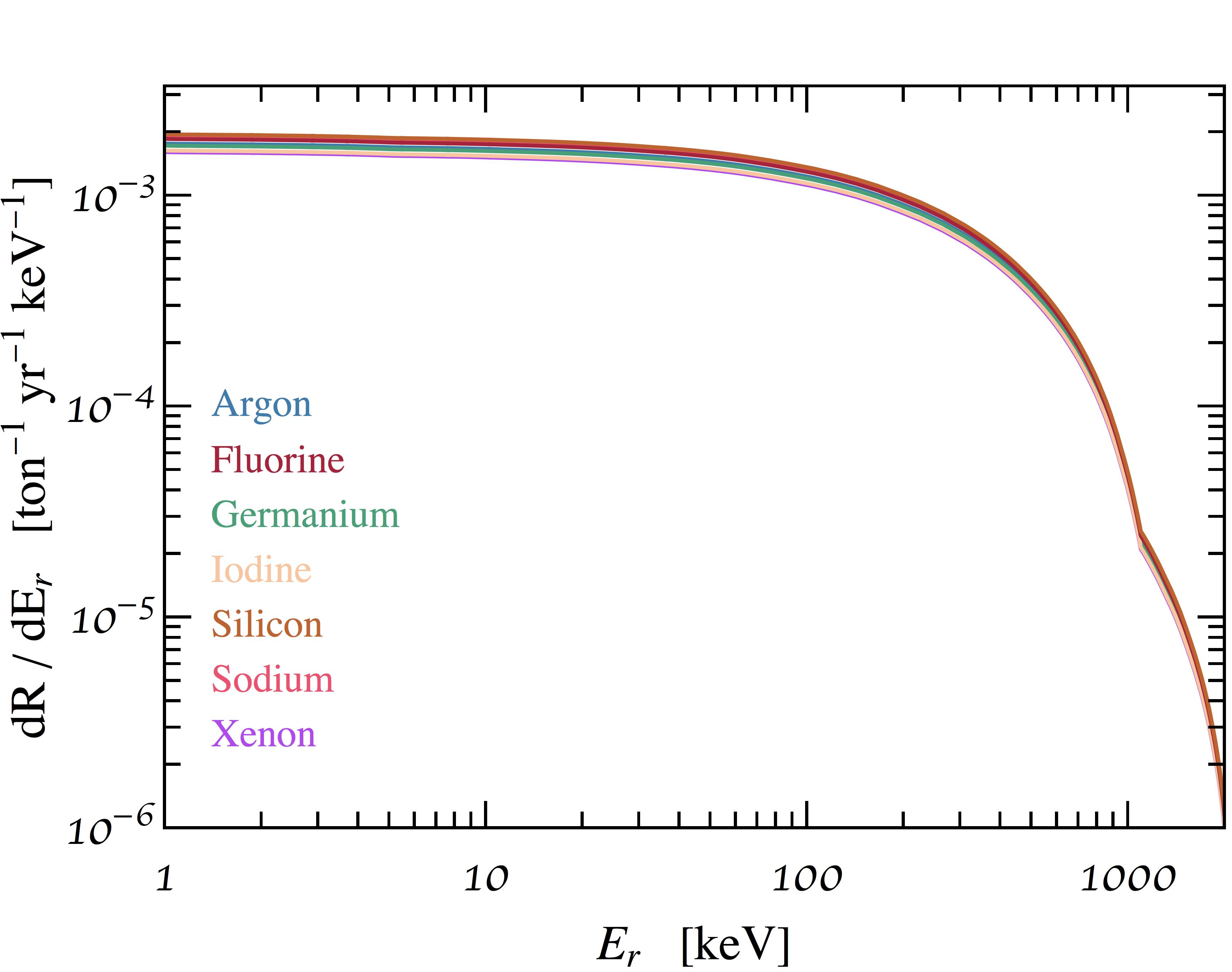} 
        \end{subfigure}
        \caption[]{ [Left] Electron recoil spectrum due to the indicated type of neutrinos for germanium. [Right] Total geoneutrino electron recoil spectra for various target materials.} 
        \label{fig:elecrec}
\end{figure*}

\subsubsection{Coherent nuclear scattering cross-section}

The Standard Model coherent-scattering neutrino-nucleus cross-section is given by~\cite{Freedman:1977xn} 
\begin{equation}\label{eq:nucxsec}
\dfrac{d \sigma^I (E_{\nu}, E_R)}{d E_R} = \dfrac{G_f^2 m_I}{4 \pi} Q_w^2 \left(1 - \dfrac{m_I E_R}{2 E_{\nu}^2}\right) F_{{\rm SI},I}^2 (E_R)~,
\end{equation}
where $m_I$ is target nuclide mass, $G_f$ is Fermi coupling constant, $F_{{\rm SI},I}(E_R)$ is the form factor, which we take this to be the Helm form factor \cite{Helm:1956zz}, $Q_w = [(1 - 4 \sin^2 \theta_{\rm W}) Z_I-N_I]$ is the weak nuclear charge, $N_I$ is the number of neutrons, $Z_I$ is the number of protons and $\theta_{\rm W}$ is the Weinberg angle. Since $\sin^2 \theta_{\rm W} = 0.223$ \cite{Patrignani:2016xqp}, the coherent neutrino-nucleus scattering cross-section follows an approximate $N_I^2$ scaling.

\begin{figure*}[htb]
        \centering
        \begin{subfigure}[b]{0.475\textwidth}
            \centering
            \includegraphics[width=\textwidth]{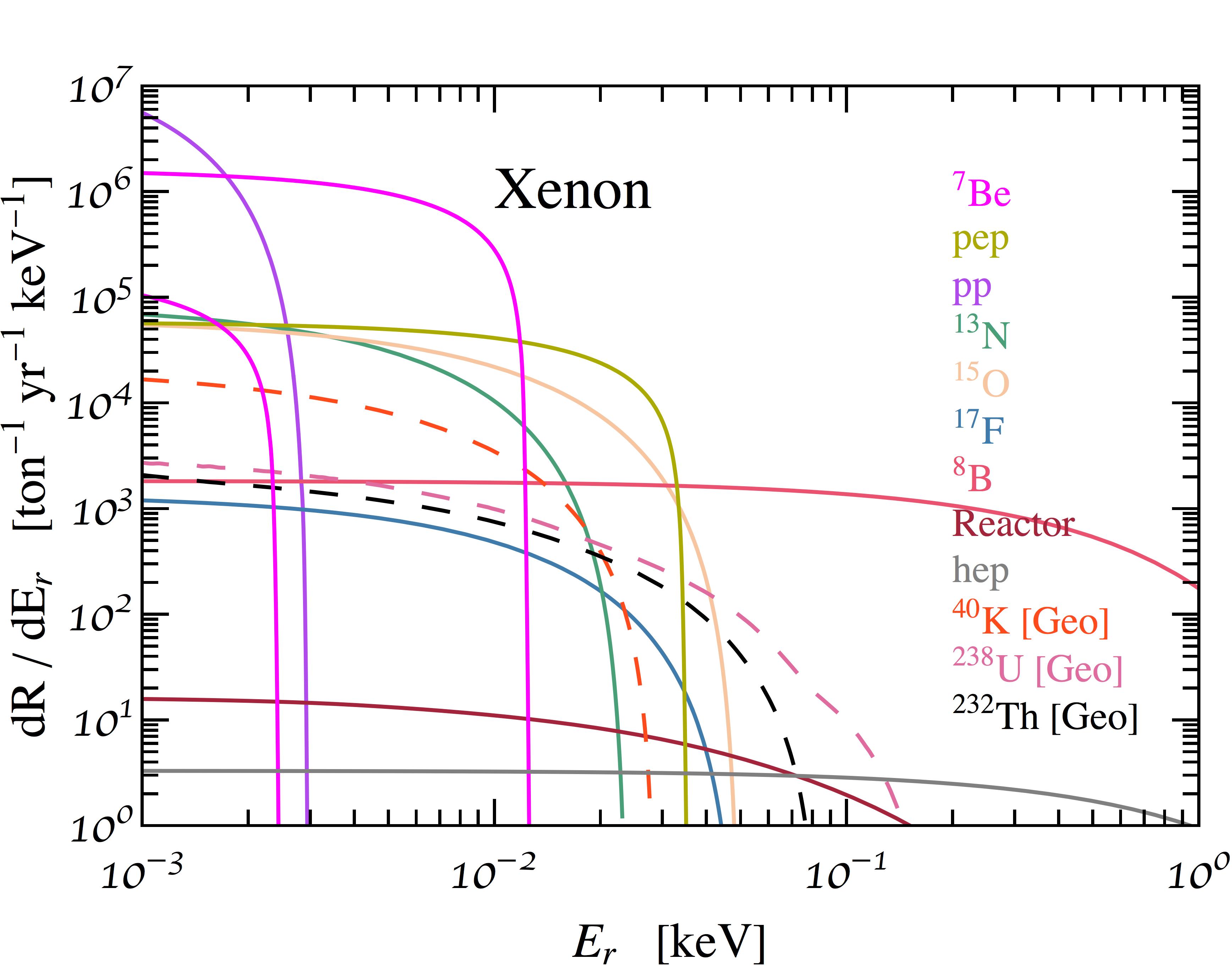} 
        \end{subfigure}
        \hfill
        \begin{subfigure}[b]{0.475\textwidth}  
            \centering 
            \includegraphics[width=\textwidth]{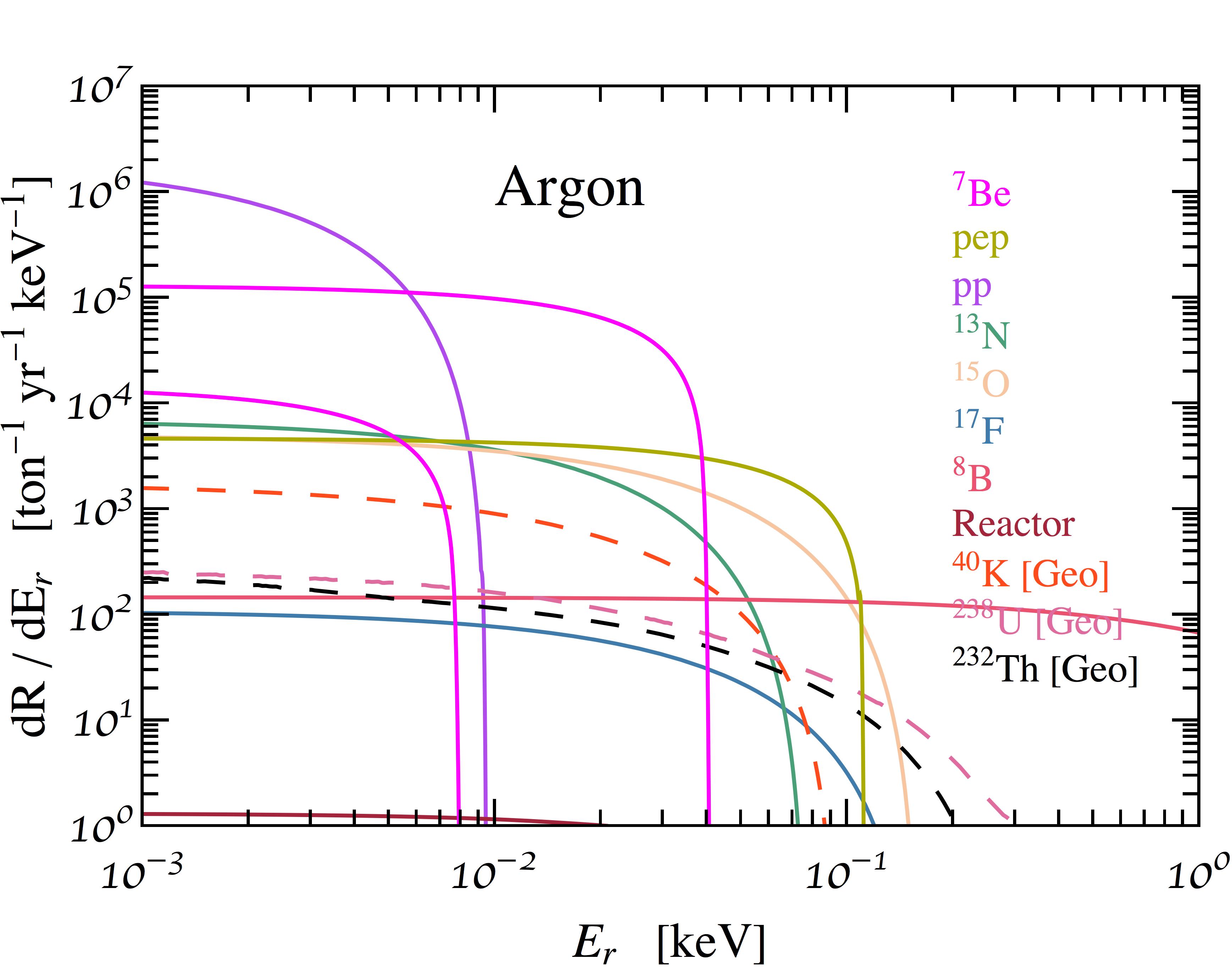} 
        \end{subfigure}
        \vskip\baselineskip
        \begin{subfigure}[b]{0.475\textwidth}   
            \centering 
            \includegraphics[width=\textwidth]{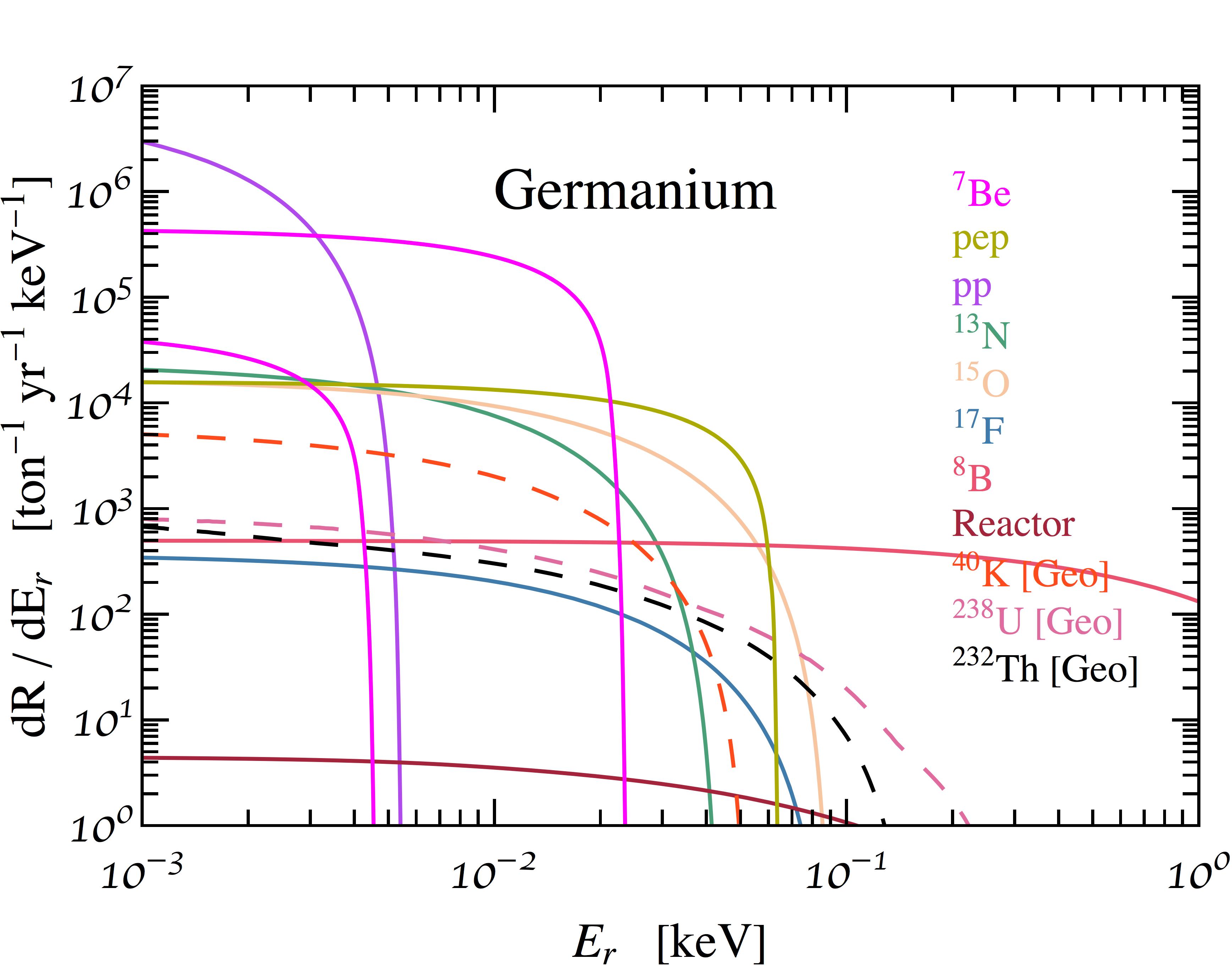} 
        \end{subfigure}
        \quad
        \begin{subfigure}[b]{0.475\textwidth}   
            \centering 
            \includegraphics[width=\textwidth]{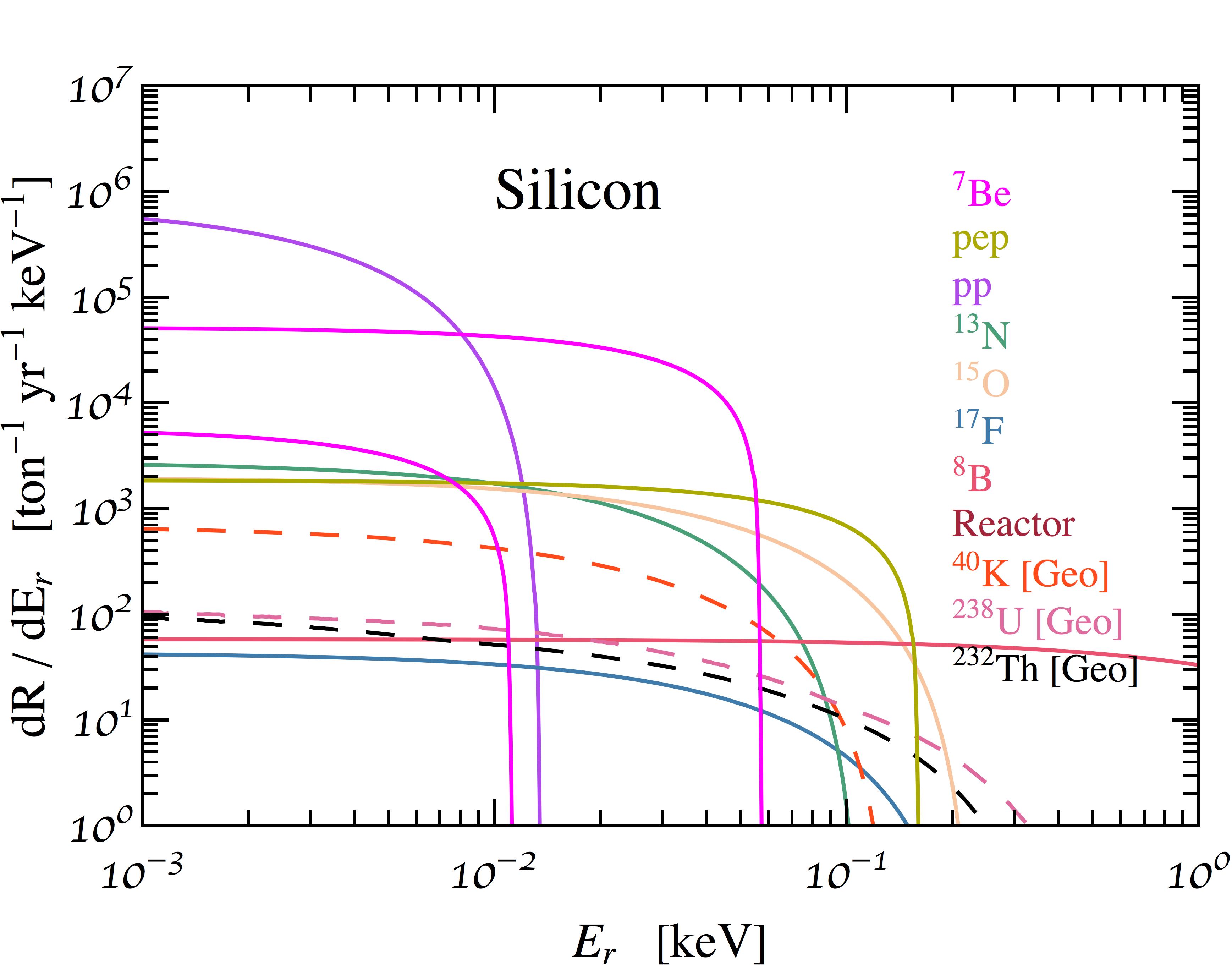} 
        \end{subfigure}
        \caption[]{Nuclear recoil spectra due to the indicated type of neutrinos for xenon, argon, germanium, silicon.} 
        \label{fig:nucrec}
\end{figure*}

\subsubsection{Elastic electron scattering cross-section}

The Standard Model neutrino-electron scattering cross-section is~\cite{Vogel:1989iv}
\begin{align}\label{eq:elecxsec}
\dfrac{d \sigma^e (E_{\nu}, E_R)}{d E_R} =& \dfrac{G_f^2 m_e}{2 \pi} \Big[g_{\nu}^2 + g_{\nu}^{\prime 2} \Big(1 - \dfrac{E_R}{E_{\nu}}\Big)^2 \notag\\ & + g_{\nu} g_{\nu}^{\prime} \dfrac{m_e E_R}{E_{\nu}^2}\Big]~.
\end{align}
Here, the flavor-dependent $g_{\nu}$ weak couplings 
\begin{equation}
g_{\nu (\mu, \tau)} = 2 \sin^2 \theta_{\rm W} - 1~,~~~g_{\nu (e)} = 2 \sin^2 \theta_{\rm W} + 1 ~
\end{equation}
are enhanced for the electron flavor, while 
\begin{equation}
g_{\nu}^{\prime}(e, \mu, \tau) = 2 \sin^2 \theta_{\rm W}~.    
\end{equation}
The corresponding anti-neutrino cross-sections are obtained via $g_{\overline{\nu} (\nu)} \leftrightarrow g_{\nu (\overline{\nu})}^{\prime}$ interchange.

The electron-neutrino cross-section of~\Eq{eq:elecxsec} is valid only for a neutrino scattering with a free electron. However, the target electrons of interest in this study are bound to nuclei. For larger recoil energies and smaller nuclei the atomic effects are expected to become negligible. On the other hand, for recoil energies of few keV in germanium~\cite{Chen:2013lba,Chen:2014dsa,Chen:2014ypv} and $\mathcal{O}(10)$ keV in xenon~\cite{Chen:2016eab}  the cross sections can become noticeably suppressed. Since we do not perform a statistical analysis for electron recoils, this effect is neglected. 

\subsubsection{Neutrino oscillations}

The coherent neutrino-nucleus scattering is mediated by the $Z$ boson. Thus, this interaction is independent of the neutrino flavor and the corresponding oscillation effects. On the other hand, electron-neutrino interactions are sensitive to   oscillation effects.

Due to the long range of production distances, the survival probability of $\overline{\nu}_e$ geoneutrinos due to averaged out oscillation effects is approximately
\begin{equation}
\langle P_{ee} \rangle = \cos^4 \theta_{\rm 13} \Big( 1 - \dfrac{\sin^2 (2 \theta_{\rm 12})}{2}\Big) + \sin^4 \theta_{\rm 13} \simeq 0.55~,    
\end{equation}
assuming $\theta_{\rm 13} = 8.5^{\circ}$ and $\theta_{\rm 12} = 34.5^{\circ}$ \cite{deSalas:2017kay}.~The Earth's Mikheyev-Smirnov-Wolfenstein
(MSW) oscillation effects for geoneutrinos are negligible~\cite{Wan:2016nhe}. 
On average, $\langle P_{ee} \rangle \simeq 0.55$ is also valid for reactor neutrinos.
For solar neutrinos we also take an electron neutrino survival probability of
$\langle P_{ee} \rangle \simeq  0.55$ for all fluxes aside $^{8}$B, since they lie in the region of vacuum oscillations of $E_{\nu} \lesssim 1$ MeV. For $^{8}$B neutrinos we take $\langle P_{ee} \rangle \simeq  0.35$, since their energies are in the matter-enhanced oscillation region of $E_{\nu} \gtrsim 5$ MeV. This choice is consistent with the large mixing angle (LMA-)MSW solution,
as indicated by  measurements~\cite{Agostini:2018uly}.

\section{Predicted recoil spectra}

\subsection{Electron recoils}

The left panel of Fig.~\ref{fig:elecrec} shows the recoil spectrum of the three types of geoneutrinos in  germanium (dashed lines). The electron recoil spectrum is almost the same for all target elements. This can be seen in the right panel of the same figure, where we show the total electron recoil spectrum (the sum of all three geoneutrino contributions) for several elements. The reason is that the ratio  $Z_I/m_I\simeq Z_I/A_I m_p$, where $m_p$ is the proton mass, which enters in the rate is approximately 1/2$m_p$ for all elements and the differential cross section Eq.~\eqref{eq:elecxsec} is the same, neglecting binding effects.

 The backgrounds for a geoneutrino signal in electron recoils due to contributions other than neutrinos are expected to be large. Thus, we do not examine detectability of geoneutrinos through electron scattering.
 
\subsection{Nuclear recoils}

\begin{figure}[htb]
\centering
\includegraphics[width=.45\textwidth]{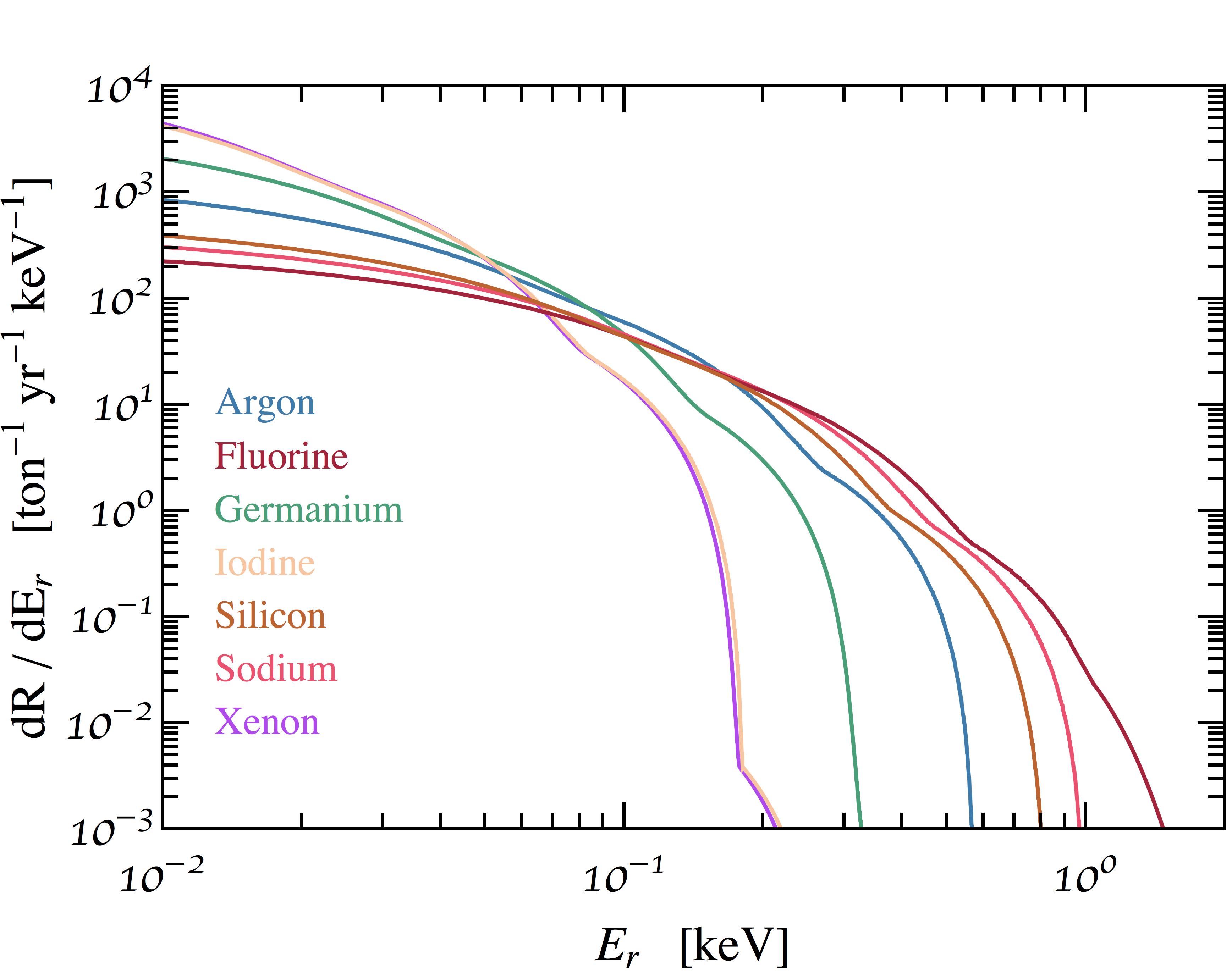}
\caption{\label{fig:geonuc_all} Total geoneutrino nuclear recoil spectra for various target materials.}
\end{figure}

Fig.~\ref{fig:nucrec} shows the three geoneutrino nuclear recoil spectra (dashed lines) for  Xe, Ar, Ge and Si. The total (sum of all three contributions) geoneutrino nuclear recoil spectra for these and other target elements are shown in Fig.~\ref{fig:geonuc_all}. These spectra, and the corresponding differential cross sections, differ considerably between various elements. With the threshold we are using (see~\Tab{tab:experiments}), Fig.~\ref{fig:nucrec} shows that $^{40}$K geoneutrino signal cannot be detected either in Xe or Ar,  and that $^{232}$Th signal cannot be observed in Xe. The lowest threshold of 0.6 keV currently employed in Xe-based experiments would not allow to detect any geoneutrino signal. The most promising targets to detect geoneutrinos are therefore Ge and Si, because of the achievable low threshold as demonstrated by the SuperCDMS SNOLAB HV (high voltage) detectors.

\section{Detection Sensitivity for Nuclear Recoils}

\subsection{Statistical analysis}

To determine the discovery sensitivity of a direct detection experiment we use a frequentist analysis based on the profile likelihood ratio test \cite{Cowan:2010js,Rolke:2004mj, Cowan:2011an}  used recently in Ref.~\cite{Gelmini:2018ogy} and other studies \cite{Billard:2011zj,Billard:2013qya,Aprile:2011hx,Ruppin:2014bra}. This test is performed by generating simulated data sets for the experiment,  assuming a normalization (i.e. an energy integrated neutrino flux) for each of the considered neutrino fluxes ${\phi}_{\nu_k}$, which we take here to be the average theoretical predicted value $\overline{\phi}_{\nu_k}$. Here $k = 1, 2, \dots , n_{\nu}$, with $n_{\nu}$ being the total number of neutrino contributions in \Tab{tab:backflux} as well as the geoneutrino fluxes in \Tab{tab:geoflux} that produce recoils above the particular experimental threshold. More specifically, for each of the ${\phi}_{\nu_k}$  parameters we generate fake data consisting of a total number of ``observed'' events $N_{\rm o}$ at particular recoil energies $E_j$ with $j = 1, 2, \dots , N_{\rm o}$ that we use to define the following likelihood function
\begin{align} \label{eq:likemain}
\mathcal{L} = &\Bigg\{\frac{e^{-N_{\rm E}}}{N_{\rm o}!}  \prod_{j=1}^{N_{\rm o}}{\rm MT} \Big(\dfrac{d R_{\rm tot}}{d E_R}\biggr|_{\ER=E_j}  \Big) \Bigg\} \notag\\ & \times \prod_{k'=1}^{n_{\nu}-n_g}\dfrac{1}{\sqrt{2 \pi}\sigma_{\nu_{k'}}}{\rm exp}\biggr[-\left(\frac{\phi_{\nu_{k'}} - \bar{\phi}_{\nu_{k'}}}{\sqrt{2}\sigma_{\nu_{k'}}}\right)^2 \biggr]~.
\end{align}
Here, $n_g$ is the number of geoneutrino species that produce recoils above threshold and $k'$ numbers the fluxes of neutrinos other than geoneutrinos, with uncertainties $\sigma_{\nu_{k'}}$ taken into account. The total differential rate is the sum of the contributions of the signal due to geoneutrinos, denoted by the index $k_g$, and the background, assumed here to be due only to all other types of neutrinos, 
\begin{equation} \label{eq:drtot}
\dfrac{d R_{\rm tot}}{d E_R} =  \sum_{k_g}^{n_g}\frac{d R_{\nu_{k_g}}}{d\ER}(\phi_{\nu_{k_g}})  + \sum_{k'}^{n_{\nu}-n_g} \frac{d R_{\nu_{k'}}}{d\ER}(\phi_{\nu_{k'}})~,
\end{equation}
defined in \Eq{eq:totnurate}, respectively. The total number of predicted events $N_{\rm E}$ is obtained by integrating \Eq{eq:drtot} over our recoil energy ROI and multiplying by the assumed exposure MT. In \Eq{eq:likemain} the extended likelihood (i.e. the term in the curly brackets) is multiplied by a Gaussian product of likelihoods, centered around the mean predicted flux normalization $\overline{\phi}_{\nu_{k'}}$, for each background neutrino species $\nu_{k'}$ ($k' = 1, 2, \dots, n_{\nu}-n_g$) to take into account the uncertainty in the $\phi_{\nu_{k'}}$  fluxes.
 In the Gaussian likelihoods the $\sigma_{\nu_{k'}}$ adopted  is the 1$\sigma$ uncertainty in the particular flux given in \Tab{tab:backflux} (see Section~\ref{ss:nuback}).
 
As explained in Ref.~\cite{Gelmini:2018ogy}, the procedure we follow to obtain each set of simulated data involves two steps: 1) find the total number of events of each type $k$ (where $k$ numbers all neutrino types above threshold), and 2) find the recoil energy for each of the fake data events. For each realization of fake data, the number of events of a specific type, $n_{k}$, is found from a Poisson distribution $P_{k}$, 
\begin{equation}
 P_{k} = \dfrac{\mu_{k}^{n_{k}} \, e^{- \mu_{k}}}{n_{k}!}~, 
\end{equation}
where the mean $\mu_{k}$ is the number of events predicted by the model being tested, defined by the particular average total neutrino flux $\bar{\phi}_{\nu_k}$, the target element, energy range and assumed exposure MT. Choosing a random number for the cumulative probability distribution of each $P_k$ one value of $n_k$ is randomly generated, according to an inverse transform sampling.
The total number $N_0$ of ``observed'' events is then $N_0 = \sum_{k = 1}^{n_{\nu}} n_k$.

To determine the energy of each one of the $n_k$ events of each neutrino type we use as probability density function (PDF) the corresponding differential recoil rate $d R_{k}/d E_{R}$ normalized by the total rate (i.e. the rate integrated over the specified energy ROI for the given experiment $R_{k}$) as
\begin{equation}
\Big(\text{PDF}\Big)_{k} = \dfrac{1}{R_{k}} \dfrac{d R_{k}}{d E_R}~.
\end{equation}
The corresponding $n_{k}$ recoil energies are again obtained with inverse transform sampling. With the above procedure, we simulate 600 data sets for each exposure and experimental configuration.

\begin{figure*}[htb]
        \centering
        \begin{subfigure}[]{0.45\textwidth}
            \centering
            \includegraphics[width=\textwidth]{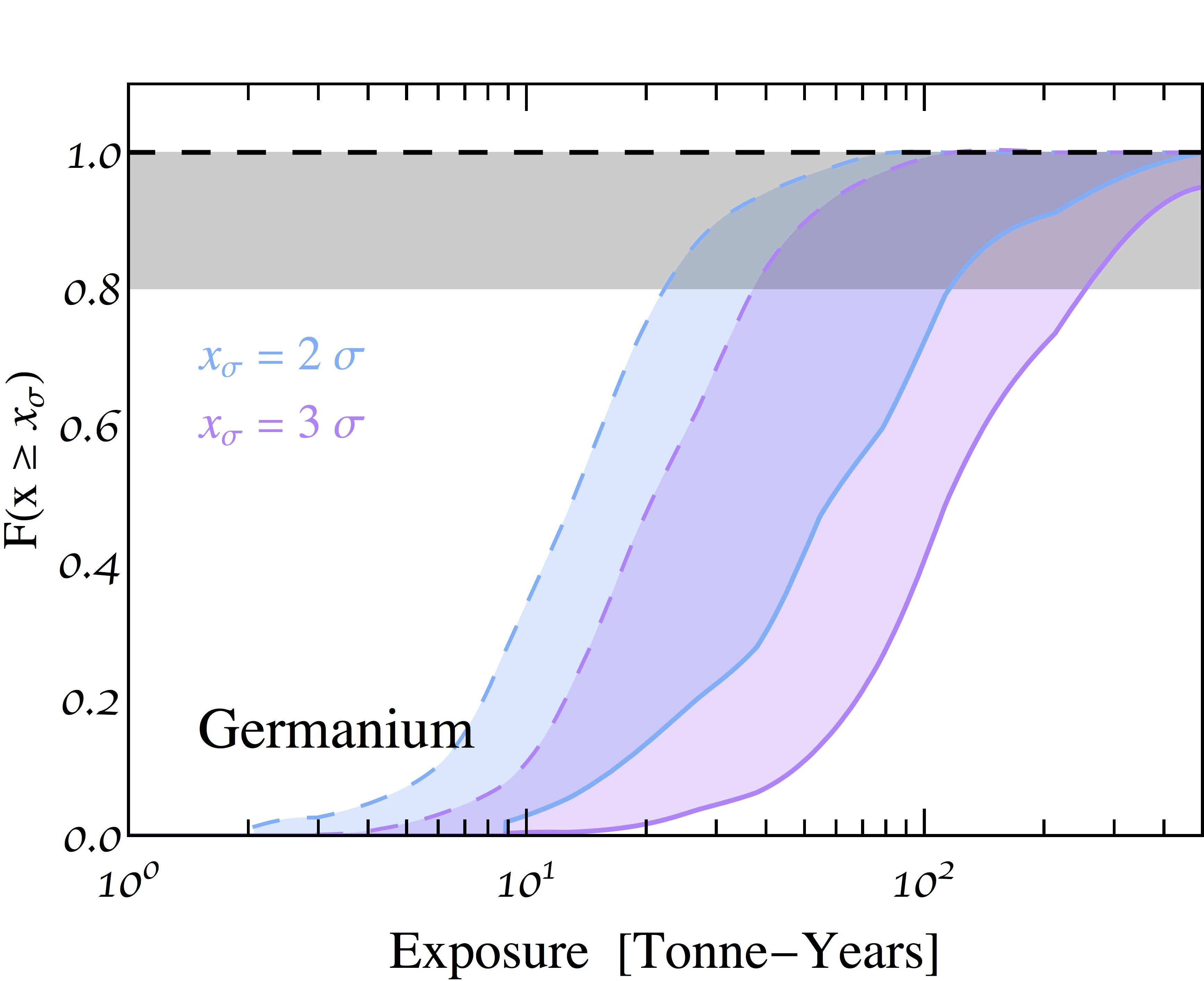} 
        \end{subfigure}
        \begin{subfigure}[]{0.45\textwidth}   
            \centering 
            \includegraphics[width=\textwidth]{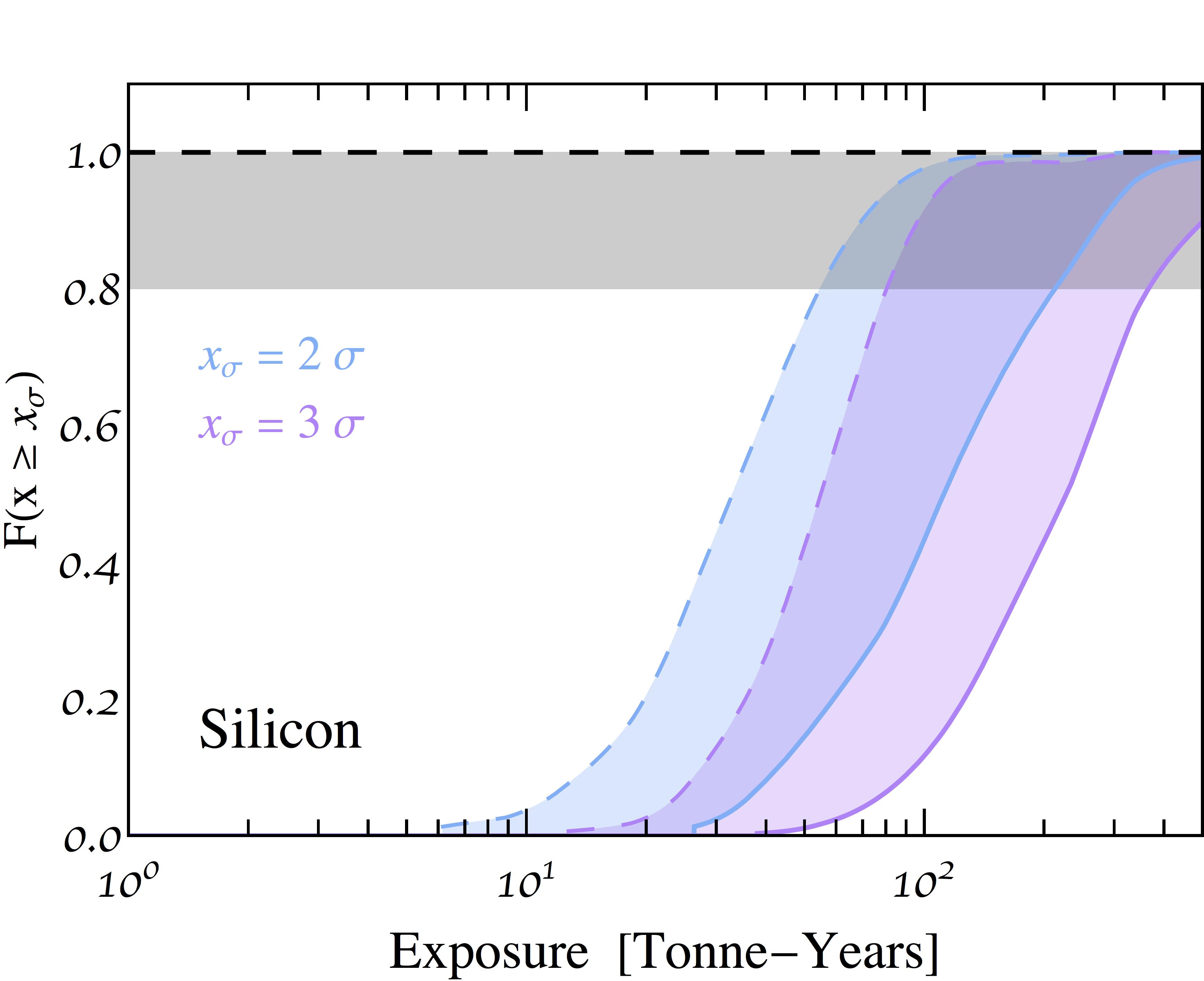} 
        \end{subfigure}
        \begin{subfigure}[]{0.45\textwidth}   
            \centering 
            \includegraphics[width=\textwidth]{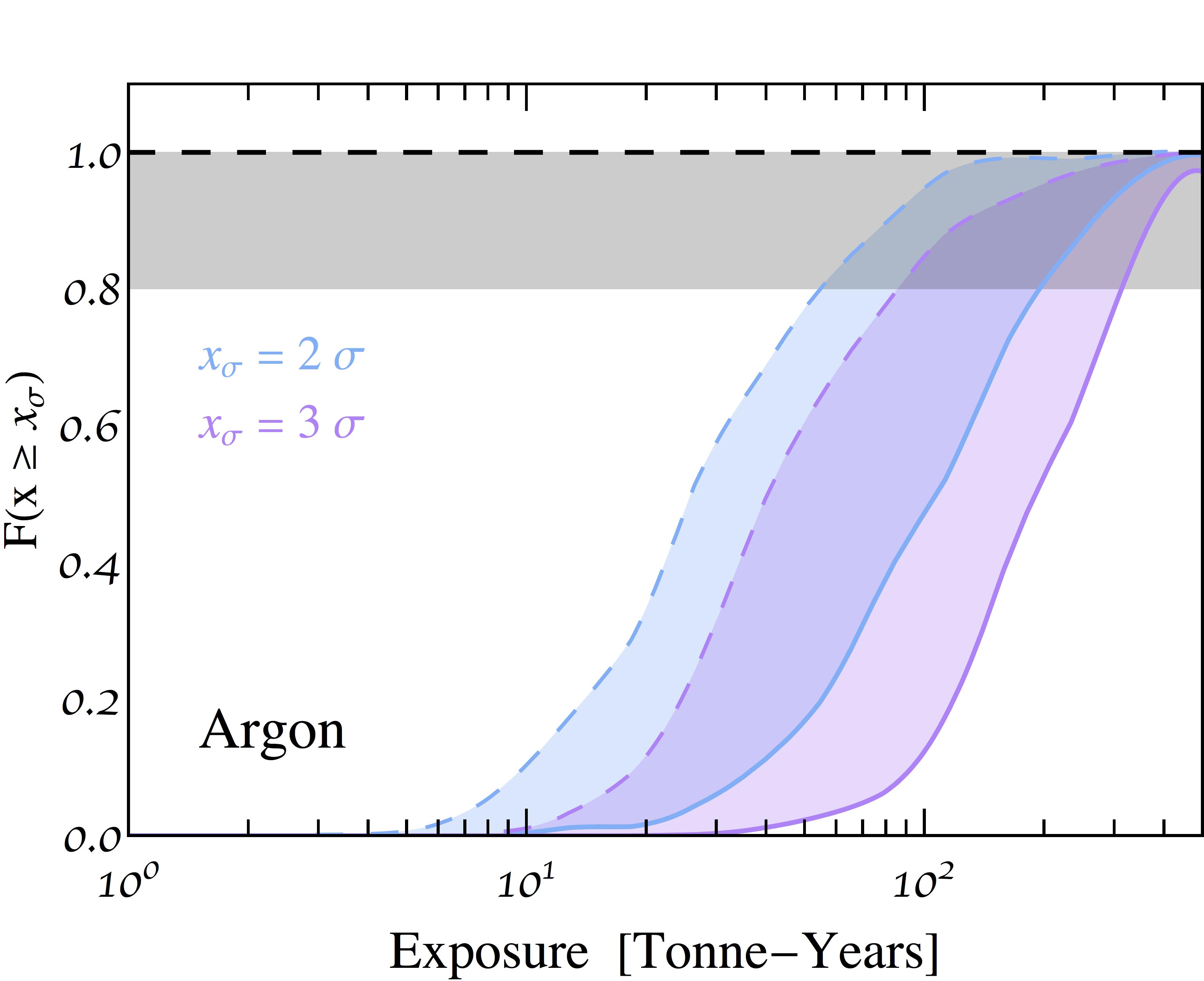} 
        \end{subfigure}
        \caption[]{Minimum exposure needed to obtain a fraction $F$ of simulated data sets in which there is a detection of geoneutrinos at the 2$\sigma$ (blue)  and 3$\sigma$ (purple) level in Ge, Si and Ar.  The solid (dashed) lines in each band assumes the current (and $\times 10^{-2}$) systematic errors of \Tab{tab:backflux}. The gray shaded region shows $F> 0.8$, where the chance of a type-2 error is small ($<0.2$).}. 
        \label{fig:nucrecsens}
\end{figure*}

For each simulated data set we define a test statistic $q_0$ that  allows one to reject the background only hypothesis $H_0$,  assuming it is true, with a probability not larger than some value $\alpha$ that denotes  the significance level of the test. In our case, $H_0$ is the assumption that the flux of geoneutrinos is zero, ${\phi}_{\nu_{k_g}} = 0$,
For our analysis we chose $\alpha$ to correspond to either 2$\sigma$, or 3$\sigma$ (i.e. $\alpha = 0.0228$ and $\alpha = 0.00135$, respectively).
 The test statistic $q_0$ for each simulated data set is the profile likelihood ratio, defined as
\begin{equation} \label{eq:likeratio}
q_0 = 
\begin{cases}
   - 2 \,\text{ln} \left( \dfrac{\mathcal{L}(\vec{\phi}_{\nu_{k_g}} = 0, \hat{\hat{\vec{\phi}}}_{\nu_{k'}})}{\mathcal{L}(\hat{\vec{\phi}}_{\nu_{k_g}}, \hat{\vec{\phi}}_{\nu_{k'}})} \right) ,& \hat{\vec{\phi}}_{\nu_{k_g}} \geq 0 \\
    0 ,& \hat{\vec{\phi}}_{\nu_{k_g}} < 0 \, ,
\end{cases}
\end{equation}
where the $\phi_{\nu_{k'}}$ are treated as nuisance parameters. 
  The hats refer to values that maximize the likelihood, with double-hats referring to values maximizing the likelihood subject to the constraint ${\phi}_{\nu_{k_g}} = 0$. Note that by definition $q_0 \geq 0$, with larger values of $q_0$ indicating greater incompatibility of the simulated data with the background only hypothesis $H_0$. Thus, we require that the probability $p_0$ of having a $q_0$ value larger (i.e. more incompatible with the data if due to background only) than the ``observed'' (i.e. than the $q_0$ of the simulated dataset) $q_0^{\rm obs}$
 \begin{equation}\label{eq:pval}
p_0 = \int_{q_0^{\rm obs}}^\infty dq_0 \, f(q_0|H_0) \, , 
\end{equation}
is not larger than $\alpha$, $p_0 \leq \alpha$. Here, $f(q_0|H_0)$ is the PDF of obtaining $q_0$ under the background-only hypothesis $H_0$. In the large sample limit, Wilks' theorem ensures that $f(q_0|H_0)$ is given by one half times a delta function at $q_0 = 0$ plus one half times a $\chi^2$ distribution with $n_g$ degrees of freedom~\cite{Cowan:2010js}, with $n_g$ corresponding to the number of geoneutrino species contributing to the signal. For xenon $n_g = 1$, argon $n_g = 2$ and for germanium and silicon $n_g = 3$. In the analyses detailed below, we require a value of $\alpha$ corresponding to a $Z\sigma$ significance with  $Z = 2$ or 3, namely $\alpha(2)= 0.0228 $ and $\alpha(3) = 0.00135$. The value of $\alpha$ corresponding to Z$\sigma$ for $n_g= 1$ (namely for 1 degree of freedom) can be obtained by requiring, approximately, that ${q_0^{\rm obs}} \geq Z^2$.  No similar approximate condition exists for $n_g= 2$ and $n_g= 2$ and the lower limit of ${q_0^{\rm obs}}$ is found numerically. 

In the Fig.~\ref{fig:nucrecsens}
 we show the fraction $F$ of simulated data sets generated with a particular exposure MT that fulfill this requirement, for $Z = 2$ or 3.
 Namely, we determine the fraction of simulated data sets that produce p-values $p_0^i \leq \alpha(Z)$ (i.e. those resulting in $\geq Z\sigma$ detection of geoneutrinos), by computing
\begin{equation}
 F({\phi}_{\nu_{k_g}}, {\rm MT}) \equiv \sum_{i=1}^{N_{\rm sim}}\frac{1}{N_{\rm sim}} \begin{cases} 1 &\mbox{if } p_0^i \leq \alpha(Z) \\ 0 &\mbox{if } p_0^i > \alpha(Z) \end{cases} \, , 
\end{equation}
 where $N_{\rm sim}$ is the number of simulated data sets with fixed parameters $({\phi}_{\nu_{k_g}}, {\rm MT})$. 

Notice that the fraction $F$ is the ``power'' of the test of $H_0$ with respect to the alternative hypothesis $H_{\sigma}$, in which ${\phi}_{\nu_{k_g}} \neq 0$ (e.g. \cite{Patrignani:2016xqp}, Sec. 40), i.e. $F= (1-\beta)$. This means that  the probability of not rejecting $H_0$ when the ``geoneutrino detection" alternative hypothesis $H_{\sigma}$ is true is less than the value $\beta= 1 - F $.

\subsection{Results}

Fig.~\ref{fig:nucrecsens} shows the two bands of fractions $F$ of simulated data sets generated with a particular exposure MT in which there is a detection of geoneutrinos at an x$\sigma$ level with x~$\geq Z$, for $Z$=2 (in blue) and $Z$=3 (in purple), as a function of the exposure MT,  for Ar, Ge and Si compositions. The higher exposure boundary (solid line) in each band assumes the systematic errors shown in \Tab{tab:backflux}. The lower exposure boundary (dashed line) of each band assumes instead a very optimistic scenario in which the errors have been divided by a factor of 100.
The gray shaded area highlights the region of $F> 0.8$, where the chance of a type-2 error is small (i.e $\beta < 0.2$).  

With a Xe target even a 2$\sigma$ detection of geoneutrinos is not reachable below a 500 tonne-year exposure, the largest we consider. The reason is that for the 0.1 keV threshold that  we assume  only a small portion of $^{238}$U geoneutrino signal is observable (see Fig.~\ref{fig:nucrec}).

Fig.~\ref{fig:nucrecsens} shows that the most promising detector configuration for geoneutrino detection is based on Ge, because of the lowest potential energy threshold.

\section{Summary}

A measurement of the level of geoneutrino emission is of utmost importance to determine the fraction of the heat emitted by Earth that remains from its formation.  The major uncertainty is due to the unknown mantle contribution to the radiogenic heat. 

Direct detection experiments have a clear advantage over scintillator-based experiments, which have been used to study geoneutrinos thus far, in that their threshold could be low enough (at least in Ge- and Si-based detectors) to observe geoneutrinos with energies below the kinematic threshold of 1.8 MeV, the lowest neutrino energy required for inverse beta decay.  This means that $^{40}$K geoneutrinos can also contribute to the signal in direct detection experiments.  A measurement of the present level of $^{40}$K with respect to the levels of other geoneutrino types  could provide insight into Earth's ``volatility curve", a central quantity for Earth's formation history~\cite{Bellini:2013wsa}. 

 In this paper we computed the electron and nuclear recoil spectra in different target materials.  In Fig.~\ref{fig:neutrino_flux} one can see that the geoneutrino electron recoil rate is about one order of magnitude lower than the rate of CNO neutrinos, whose detection sensitivity has been recently  explored~\cite{Billard:2014yka,Cerdeno:2017xxl,Newstead:2018muu}.
 
 For the detection of nuclear recoils from geoneutrino signal we considered several potential experimental setups with optimistic thresholds and located in the Jinping Underground Laboratory in China, where the expected geoneutrino flux is significantly larger than in other
 laboratory sites, because of its proximity to the Himalayas, where the crust is the thickest. Assuming that radiogenic heat accounts for most of the heat emitted by our planet, a viable possibility, we  study the minimal required exposures for a geoneutrino observation at the 2 and 3$\sigma$ level, if the background is due only to other neutrinos (solar and reactor neutrinos).
 
 The results of our analysis are shown in Fig.~\ref{fig:nucrecsens}. The most promising target material analyzed in this study for geoneutrino detection is germanium, which could have the lowest experimental threshold, if the exposure reaches $\mathcal{O}(10)$ tonne-years.

\section*{Acknowledgements}

We thank Carolina Lithgow-Bertelloni, Lars Stixrude, Gerald Schubert and Linyan Wan for discussions and Sergio Palomares-Ruiz for comments. The work of GBG and VT was supported in part by the
U.S. Department of Energy (DOE) under Grant No. DE-SC0009937. SJW acknowledges support
from Spanish Ministry of Economy and Competitiveness (MINECO) national grants FPA2014-57816-P and FPA2017-
85985-P, and from the European projects H2020-MSCAITN-2015//674896-ELUSIVES
and H2020-MSCA-RISE2015.

\bibliography{geonulib}
\end{document}